\documentclass[unnumsec,webpdf,contemporary,large]{oup-authoring-template}%
\graphicspath{{fig/}}

\theoremstyle{thmstyleone}%
\theoremstyle{thmstyletwo}%
\theoremstyle{thmstylethree}%

\usepackage[normalem]{ulem}
\usepackage{bm,bbm}
\usepackage{pifont}
\newcommand{\cmark}{\ding{51}}%
\newcommand{\xmark}{\ding{55}}%

\begin{document}

\journaltitle{}%Journal title}
\DOI{}%https://doi.org/10.xxxx/xxx}
\copyrightyear{}%2024}
\pubyear{}%2024}
%\access{Advance Access Publication Date: Day Month Year}
\appnotes{Review}

\firstpage{1}

\subtitle{Genetic and population analysis}

\title[Survive with omics]{Tutorial on survival modeling with applications to omics data}

\author[1,2,$\ast$]{Zhi Zhao\ORCID{0000-0003-2325-1438}}
\author[1,2]{John Zobolas\ORCID{0000-0002-3609-8674}}
\author[1,3,\dag]{Manuela Zucknick\ORCID{0000-0003-1317-7422}}
\author[1,2,4,\dag]{Tero Aittokallio\ORCID{0000-0002-0886-9769}}

\authormark{Zhao et al.}

\address[1]{Oslo Centre for Biostatistics and Epidemiology (OCBE), Department of Biostatistics, Faculty of Medicine, University of Oslo, Norway}
\address[2]{Department of Cancer Genetics, Institute for Cancer Research, Oslo University Hospital, Norway}
\address[3]{Oslo Centre for Biostatistics and Epidemiology (OCBE), Research Support Services, Oslo University Hospital, Norway}
\address[4]{Institute for Molecular Medicine Finland (FIMM), HiLIFE, University of Helsinki, Finland}

\corresp[$\ast$]{Corresponding author: \href{email:zhi.zhao@medisin.uio.no}{zhi.zhao@medisin.uio.no}\ \ 
\dag Equal contribution
}

%\received{Date}{0}{Year}
%\revised{Date}{0}{Year}
%\accepted{Date}{0}{Year}

\editor{Associate Editor: Name}

\abstract{
  {\bf Motivation:} Identification of genomic, molecular and clinical markers prognostic of patient survival is important for developing personalized disease prevention, diagnostic and treatment approaches. 
  Modern omics technologies have made it possible to investigate the prognostic impact of markers at multiple molecular levels, including genomics, epigenomics, transcriptomics, proteomics and metabolomics, and how these potential risk factors complement clinical characterization of patient outcomes for survival prognosis. 
  However, the massive sizes of the omics data sets, along with their correlation structures, pose challenges for studying relationships between the molecular information and patients' survival outcomes. \\
  {\bf Results:} We present a general workflow for survival analysis that is applicable to high-dimensional omics data as inputs when identifying survival-associated features and validating survival models. 
  In particular, we focus on the commonly used Cox-type penalized regressions and hierarchical Bayesian models for feature selection in survival analysis, which are are especially useful for high-dimensional data, but the framework is applicable more generally. \\
  {\bf Availability and implementation:} A step-by-step R tutorial using The Cancer Genome Atlas survival and omics data for the execution and evaluation of survival models has been made available at \url{https://ocbe-uio.github.io/survomics/survomics.html}.\\
  %{\bf Contact:} zhi.zhao@medisin.uio.no
}

\maketitle

\section{Introduction}

Personalized medicine improves patient diagnosis and treatment by making use of patient-specific genomic and molecular markers that are indicative of disease development. 
Time to an event of interest (for example, time to death or disease progression) is a widely-used end point and patient outcome for many diseases, and therefore it has become popular to identify genomic, molecular and clinical markers for survival or progression prediction of patients suffering from complex diseases such as cancer. 
In this tutorial, we will only consider so-called right-censored survival data, where a patient has been followed for a certain time period and the event of interest is either observed in this time period or might occur at a later (as yet unobserved) time point.
Right-censored survival data include both a time and the status of each patient at that time as joint outcomes, where time is a continuous variable and status is a binary variable indicating whether the event of interest has been observed up to the given time or not; in the latter case, we refer to this observation as censored at the observed time. 
See \hyperref[tab:survival]{Table~\ref*{tab:survival}} for an example illustration, and more details of the survival data in Section ``\hyperref[sec:data]{Data}''. 
When using the status label as an outcome in an ordinary logistic regression, the regression coefficients will become increasingly uncertain and less reliable with increasing follow-up time \citep{Green1983}. 
When using the observed time (or its transformation, e.g., logarithm of time) as an outcome in an ordinary linear regression, the presence of censored observations (i.e., patients still alive by the end of follow-up period) causes considerable difficulties for assessing the accuracy of predictions \citep{Henderson1995}. 
For example, \hyperref[tab:survival]{Table~\ref*{tab:survival}} shows an example where the proportion of patients surviving past 10 years is $1/5 = 20\%$ based on the observed data, but the (unobserved) factual proportion surviving past 10 years is $3/5 = 60\%$. 

\begin{table}[!h]
  \caption{\it An example of right-censored time-to-event (survival) outcomes, illustrated in an example where the outcome of interest is the time from disease diagnosis to death. If a patient was alive at the last observed time point, the measurement is censored at that time point. \label{tab:survival}}
  {\tiny
  \begin{tabular}{cccc}
  \toprule
  {\bf Patient} & {\bf Observed time} & {\bf Status at} & {\bf Factual time to death,} \\
  ID & \bf{(years)}& \bf{observed time}& \bf{possibly unobserved (years)}\\
  \midrule
  1001      & 11   & censored & 20 \\
  1002      & 4   & dead & 4 \\
  1003      & 5   & censored & 12 \\
  1004      & 9   & dead & 9  \\
  1005      & 1   & censored & 11  \\
  \botrule
  \end{tabular}
  }
\end{table}

The availability of multiple types of genomic and molecular data poses great opportunities but also further challenges for building effective statistical models to identify biomarkers that are prognostic of patient survival. 
For example, omics profiles, such as those from mRNA expression, DNA copy number, single-point and other genetic mutations, may be available from the same patient, and these high-dimensional data come with intra- and inter-dataset correlations, heterogeneous measurement scales, missing values, technical variability and other background noise \citep{Hasin2017}. 
\cite{Rahnenfuhrer2023} provided a general guideline for high-dimensional omics and electronic health records data analysis, and discussed statistical challenges and opportunities for survival modeling. 
\cite{Bovelstad2007} and \cite{Bovelstad2009} compared various machine learning methods and showed improved survival prediction performance by coefficient shrinkage methods that combine several data sources, in their case clinical and gene expression data. 
Another recent study by \cite{Bommert2022} performed a benchmark of filter methods for feature selection in high-dimensional gene expression data for survival prediction, and recommended using the simple variance filter before fitting a $\ell_2$-regularized Cox regression model \citep{Simon2011} for accurate survival prediction and stable feature selection. 
\cite{Vinga2021} reviewed structured penalized regressions for analyzing high-dimensional omics data for survival prediction or evaluation. 
Multiple studies have shown that it is possible to further improve the prediction accuracy and feature selection by considering more complex structures, such as biological pathways, or by identifying significant features among functional relationships between the omics features \citep{Chekouo2015, Kundu2018, Wang2020, Madjar2021}. 

In this tutorial, we describe a general workflow for survival analysis with omics data, as well as review the commonly used statistical methods for feature selection and survival prediction; importantly, we provide a step-by-step R tutorial using publicly available omics data from The Cancer Genome Atlas (TCGA) project (\url{http://cancergenome.nih.gov}). 
In this example dataset, the overall survival time, demographic and gene expression data from primary invasive breast cancer patients in TCGA (TCGA-BRCA) were retrieved from the Genomic Data Commons Data Portal data release 32.0-36.0. 
Compared to the previous reviews and benchmarks of survival models in bioinformatic applications, this tutorial provides a complete workflow ranging from data preparation to final calibrated models, with a particular focus on building survival models using high-dimensional omics data, and as such covers both the commonly used penalized regressions and Bayesian models for survival analysis with high-dimensional and generally noisy datasets. 
For this tutorial, we assume that readers have the knowledge of basic statistical methods. 
Terms beyond basic statistics are explained in corresponding text.

\section{Data categories}\label{sec:data}

We use omics data and overall survival outcomes from cancer patients as an example in this tutorial, but the methods are applicable also to other diseases with similar data types in personalized medicine applications. 
The ultimate goal of personalized medicine is to identify patient-specific risk factors to guide disease prevention, diagnostic and treatment strategies. 
The identification of potential risk factors for cancer patients often considers clinical, demographic, genomic and molecular information, and their associations with the patients' time-to-event data (i.e., survival).

\begin{figure}[!h]
\centering
\includegraphics[height=0.24\textwidth]{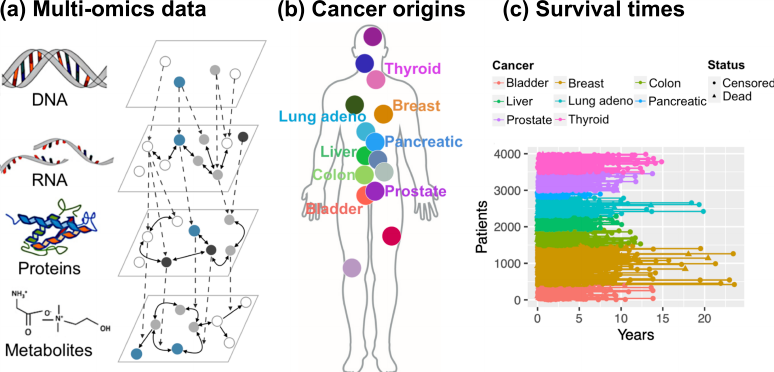}
\caption{\it Pan-cancer survival data and omics signatures. (a) Multiple omics layers (modified from \cite{Haukaas2017} and \cite{Jendoubi2021}). The network illustrates intra- and cross-layered biological features or molecules (e.g., DNA methylation, mRNA, proteins, metabolites). (b) Different origins of tumors represented by the TCGA pan-cancer patient cohorts. (c) Overall survival times of cancer patients from TCGA. Novel methods are needed to model the high-dimensional multi-omics data and leverage information from heterogeneous cancer cohorts for improved survival prognosis and biomarker identification.} \label{fig:omics}
\end{figure}

\subsection{Time-to-event data}\label{sec:datasurvival}

Time-to-event or survival data contain the event of interest (e.g., death is the event assumed in this section), together with the time from the beginning of a study period either to the occurrence of the event, end of the study, or patient loss to follow-up (i.e., right censoring; discussions of right, left and interval censoring can be found in \citealp{Leung1997}).
\hyperref[fig:omics]{Figure~\ref*{fig:omics}c} shows the survival or right-censored times since cancer diagnosis of patients. 
The exact survival time of a patient may be not observed due to censoring. 
Therefore, a patient has two outcome indices: censoring indicator $\delta$ (also called status) and observed time $\tilde T=\min\{T^*, C\}$, where $T^*$ is the exact survival time and $C$ is the censoring time. 
Indicator $\delta$ can also be denoted as $\delta=\mathbbm 1 \{T^*\le C\}$, where $\mathbbm 1 \{\cdot\}$ is an indicator function. 
To characterize the survival time of a patient, we can use survival function 
$S(t) = \mathbbm{P}\{T^*>t\},$ 
which gives the probability of the patient's survival beyond time $t$. 
Another useful quantity is the hazard function
\begin{equation*}
h(t) = \lim_{\Delta t \to 0} \frac{\mathbbm{P}\{t \le T^* <t+\Delta t| T^* \ge t\}}{\Delta t},
\end{equation*}
which is the instantaneous probability of the patient's death at time $t$ conditional on the patient having survived up to that time point. 

\subsection{Clinical and demographic data}

There are multiple sources of patient-level information that can be explored to identify risk factors for cancer patients. 
One can start with routinely collected and commonly used patient data, such as clinical and demographic variables. For example, an older male patient with a low body mass index (BMI) has a relatively high risk of gastric cancer \citep{Nam2022}. 
\hyperref[tab:clinical]{Table~\ref*{tab:clinical}} illustrates selected clinical and demographic variables often available for cancer patients. 
Clinical and demographic variables are considered important sources of information for predicting survival and are often used to build reference models for omics-based prognostic models \citep{Herrmann2021}. 

\begin{table}[!h]
  \caption{\it Examples of clinical and demographic variables\label{tab:clinical}}
  {\tiny
  \begin{tabular}{l l}  
  \toprule
  {\bf Variable} & {\bf Data type} \\
  \midrule
  Sex      & binary   \\
  Body mass index (BMI, $kg/m^2$) & continuous \\
  Ethnicity    & nominal   \\
  Age at first diagnosis in years      & integer   \\ 
  Pathological stage & ordinal   \\ 
  Therapy type (e.g., chemo-, hormone, immuno-therapy) & nominal   \\ 
  \botrule
  \end{tabular}
  }
\end{table}

\subsection{Omics data}\label{sec:datamultiomics}

Thanks to the rapid development of modern molecular biotechnology, large amounts of human genomic and molecular data have become available from many patient profiling projects. 
These projects often collect multiple levels of molecular information such as genomics data for DNA variation, transcriptomics data for mRNA expression, proteomics data for protein abundance and metabolomics data for metabolite processes, as illustrated in \hyperref[fig:omics]{Figure~\ref*{fig:omics}a}. 
Among the multiple omics levels, metabolomics is the closest to observable phenotypes, such as tumor growth and proliferation \citep{Cairns2011}. 
To deeply understand the molecular biology of tumor development, multiple levels of omics data may deliver novel insights into the circuits of molecular interactions that underlie the disease initiation and progression Tarazona et al. (2021).

DNA-level omics data often include single nucleotide polymorphisms (SNP), DNA methylation, and somatic copy number variation, see illustration of these data types in \hyperref[tab:DNA]{Table~\ref*{tab:DNA}}. 
Each SNP feature can be coded as $\{0,1,2\}$, according to the number of minor alleles at the given locus in each individual, i.e., AA, AG, GC or TT. 
DNA methylation reveals methyl groups added to the DNA molecule, which can be quantified as a $\beta$-value $\beta = \frac{M}{M+U+a}\in [0,1]$, where $M$ and $U$ are the fluorescence intensities of the methylated and unmethylated DNA at a CpG locus, and the offset $a$ is often set to 100 to stabilize the $\beta$-value when $M$ and $U$ are small. 
Somatic copy number variation measures the number of repeated sections of the tumor genome. 
\hyperref[tab:DNA]{Table~\ref*{tab:DNA}} shows the assumed distributions for the downstream statistical modeling. 
For example, DNA methylation $\beta$-value is a proportion, often either close to 1 or 0, which can be characterized by a mixture of two beta distributions. 

\begin{table*}[t]
  \caption{\it \centering{Illustration of omics data and distribution assumptions for data generated by commonly used high-throughput technologies.} \label{tab:DNA}}
  {\tiny
\makebox[\textwidth][c]{
  \begin{tabular}{cccccccc}
  \toprule
  {\bf Patient ID} & {\bf SNP} & {\bf Methylation } & {\bf Copy number variation} & {\bf Gene expression } & {\bf miRNA } & {\bf Protein } & {\bf Metabolite } \\
 &  & {\bf ($\beta$-value)} & {\bf (number of copies)} & {\bf (reads)} & {\bf (reads)} & {\bf (intensity/concentration)} & {\bf (intensity/concentration)} \\
  \midrule
  1001      & 1   & 0.2  & 0 & 5 & 5 & 0.07 & 0.07\\
  1002      & 0   & 0.11 & 5 & 2 & 2 & 0.1 & 0.1\\
  1003      & 1   & 0.95 & 10 & 0 & 0 & 9.6 & 9.6\\
  1004      & 2   & 0.5  & 4  & 30 & 30 & 2.8 & 2.8\\
  $\vdots$ &  $\vdots$ &  $\vdots$ &  $\vdots$ &  $\vdots$ &  $\vdots$ &  $\vdots$  &  $\vdots$ 
  \medskip\\
  {\bf Technology} & SNP array  & Infinium BeadChip  & SNP array (Gistic2) & RNA-Seq & RNA-Seq & Mass spectrometry & Mass spectrometry
  \smallskip\\
  {\bf Distribution} & Ordinal & Beta mixture & Negative binomial & Negative binomial & Negative binomial & log-Gaussian & log-Gaussian \\
  {\bf assumptions}  &  &  & log$_2$ ratio: Gaussian & log$_2$ scale: Gaussian & log$_2$ scale: Gaussian & log$_2$ scale: Gaussian & log$_2$ scale: Gaussian \\ 
  \botrule
  \end{tabular}}
  }
\end{table*}

RNA-level omics data usually include messenger RNA (mRNA) expression and microRNA (miRNA) expression. 
Traditionally, DNA microarrays were used to measure the expression levels of DNA sequences called probes, which acted as a proxy for the amount of reads representing a genomic feature of interest. 
The reads of microarray expression level can be characterized by a negative binomial distribution directly, or a Gaussian distribution after log$_2$ transformation (see \hyperref[tab:DNA]{Table~\ref*{tab:DNA}}). 
Nowadays, RNA sequencing (RNA-seq) has replaced DNA microarrays, since it allows for the sequencing of the whole transcriptome, while DNA microarrays only profile predefined transcripts or gene probes. 
miRNAs are small noncoding regulatory RNAs that play an important role in regulating gene expression and are highly evolutionary conserved in mammals \citep{Bartel2018}. 

Protein-level omics data usually originates from mass spectrometry (MS)-based proteomics profiling of a particular cell, tissue, organ, which can detect and measure abundance levels of entire or phosphorylated proteins. 
In contrast to global proteomics with MS, TCGA consortium has produced more targeted proteomics profiles, involving a set of 180-250 protein features using reverse-phase protein arrays (RPPA) \citep{Akbani2014}. 
In contrast, The Clinical Proteomic Tumor Analysis Consortium (CPTAC) profiling has produced more than 10000 protein features using MS technology \citep{Edwards2015}. 
Protein expression data can often be considered approximately Gaussian distributed after log2 transformation, depending on the data-generating platform. 
In global proteomics, there are often sample-specific missing values due to detection limits of protein quantification.

Metabolite-level omics data has similar statistical properties to proteomics data, and metabolomics profiling is usually done with MS-based technologies, enabling the detection and quantification of many thousands of metabolic features simultaneously. 
Nuclear magnetic resonance (NMR) spectroscopy is the other main analytical technology to profile metabolic processes. 
An important limitation of NMR spectroscopy is its relatively low sensitivity, which may lead to relatively few detected metabolites. 
Metabolic concentrations often follow logarithmic Gaussian distribution. 
Similar to large-scale proteomics, missing values due to detection limits should be treated differently than missing values due to measurement artefacts, which are more frequent in metabolite-level omics data \citep{Sun2023}.

Single-cell sequencing is becoming increasingly more prevalent in many profiling studies. 
The modern single-cell omics technologies can produce multiple levels of omics data derived from the same samples, such as transcriptomics and chromatin accessibility. 
The single-cell data types are similar to the illustration in \hyperref[tab:DNA]{Table~\ref*{tab:DNA}}, but each measurement originates from the level of individual cells. 
This article focuses mainly on survival analysis with single bulk omics data. 

\subsection{Missing data}

Missing values are often observed in many types of high-dimensional omics data due to various experimental reasons \citep{Aittokallio2009, KongW2022}. 
For example, mRNA transcriptomics data from microarrays have 1\%-10\% missing values affecting up to 95\% of genes due to corruption of image, scratches on the slides, poor hybridization, inadequate resolution, fabrication errors \citep{deBrevern2004,Tuikkala2006}; MS-based metabolomics data have 10\%–20\% missing values affecting up to 80\% of variables due to lack of peak identification by chromatogram, limitation of computational detection, measurement error, and deconvolution errors \citep{Hrydziuszko2012,Sun2023}.
The aforementioned technical reasons can lead to missing data that is either missing at random (MAR) or missing not at random (MNAR). 
When dealing with MNAR data, traditional imputation methods like multiple imputation may introduce bias. 
It is thus recommended to remove omics features which have large proportion (e.g., 50\%) missingness over patients, and then apply imputation methods (e.g., $k$-nearest neighbor imputation) for the rest of the features with missing values before doing any statistical analysis or modeling. 
Alternatively, imputation-free methods (e.g., mixture models) that can deal with missing values can be applied directly \citep{Taylor2022}. Single-cell RNA-seq (scRNA-seq) data has a vast number of zeros, so-called gene dropout events, leading highly scarce data matrices.  
\cite{Jiang2022} discussed the sources of biological and non-biological zeros in scRNA-seq data and the existing approaches for handling them.

\section{Survival analysis with low-dimensional input data}

Let us assume we have data $\mathcal D=\{(\tilde T_i, \delta_i, \mathbf X_i): i=1,\cdots,n\}$ for $n$ patients, where $\tilde T_i$ is the observed survival time, $\delta_i$ the censoring indicator and $\mathbf X_i$ contains $p$ covariates including clinical, demographic and omics features. 
To estimate a survival function $S(t)$ given the data $\mathcal D$, one needs to keep track both of the number of patients at risk and those who left the study at any time point (here we only consider the case of right censoring and assume no delayed entry). 
At time $t$ there are $Y(t)=\sum_{i=1}^n Y_i(t)$ patients at risk, where $Y_i(t) = \mathbbm 1\{\tilde T_i \ge t\}$ is the indicator that patient $i$ is at risk. 
The non-parametric Kaplan-Meier (KM) estimator \citep{Kaplan1958} uses the multiplication rule for conditional probabilities to obtain an estimation of the survival function
$\hat S(t) = \prod_{\substack{k=1\\ T_k\le t}}^K \left\{1 - \frac{d_k}{Y(T_k)} \right\},$ 
where all events happen (e.g., patients die) at $K$ distinct times, $T_1 < T_2 < \cdots < T_K$ $(K\le n)$ and there are $d_k \geq 1$ events happened at the time $T_k$. 
If no two events happen at the same time point, $d_k=1$ and $k=1,\cdots,n$. 
The KM estimator gives an estimate of the marginal survival function, i.e., when you disregard the information from the covariates. 

\begin{figure}[!h]
  \centering
  \includegraphics[height=0.26 \textwidth]{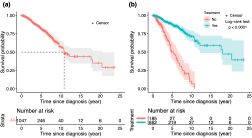}
  \caption{\it Kaplan-Meier curves of TCGA-BRCA data. (a) KM curve of all the TCGA-BRCA patients' survival data. (b) KM curves of the TCGA-BRCA patients' survival data grouped by treatment (i.e., pharmaceutical/radiation therapy) or nontreatment. The log-rank test is used to compare the two survival distributions corresponding to the two groups of patients.}\label{fig:KM}
\end{figure}

\hyperref[fig:KM]{Figure~\ref*{fig:KM}a} shows the KM curve for TCGA-BRCA primary breast cancer patients data. 
Some basic statistics can be revealed from the survival curve. For example, the estimated median survival time, i.e., the time when the survival probability is 50\%, of all the breast cancer patients is 10.8 years (dashed line in \hyperref[fig:KM]{Figure~\ref*{fig:KM}a}), and 1-, 5- and 10-year survival probabilities are 0.988, 0.853 and 0.658, respectively. 
A log-rank test \citep{Peto1972} can be used to test whether two groups of patients (e.g., with treatment (pharmaceutical/radiation therapy) or nontreatment) have the same (null hypothesis) or different survival functions (alternative hypothesis), and provide a corresponding $p$-value (see \hyperref[fig:KM]{Figure~\ref*{fig:KM}b}).
The log-rank test can also be used to compare the survival probabilities of any subgroups of patients based on other categorical variables. 

In the case where multiple clinical, demographic or omics features are available, one can explore each variable's association with survival outcomes separately. 
For a categorical variable, KM curves and log-rank tests can be used to investigate whether there is a difference between multiple survival curves categorized by the variable. 
For a continuous variable $X$, the semi-parametric Cox proportional hazards model (Cox model, \citealp{Cox1972}) is often used:
\begin{equation*}
h(t | X) = h_0(t) \exp\{X\beta\},\tag{1}\label{eq:Cox}
\end{equation*}
where $h_0(t)$ is the baseline hazard function and is left unspecified. As the name ``proportional hazards'' implies, the Cox-model estimated hazard functions of two individuals (with different values of the covariate $X$) are indeed proportional, because $h_0(t)$ does not depend on $X$ and is thus assumed to be the same for all individuals. 
The functional form \eqref{eq:Cox} describes the log-linear relationship between variable $X$ and the hazard $h(t | X)$ at any given timepoint $t$. 
It may be difficult to satisfy the log-linear relationship based on the original scale of some omics features, e.g., gene expression data from a DNA microarray study; in those cases, the use of $\log_2$-transformation of the data (\hyperref[tab:DNA]{Table~\ref*{tab:DNA}}) can be helpful.
The Cox model can also be used to investigate the risk of a categorical variable, provided the above assumptions are satisfied, which provides an effect estimate (hazard ratio, HR), in addition to the associated $p$-value. 

It is often of interest to gain insights into the multiple factors and their cooperation for the survival outcomes. 
Multivariable statistical analysis plays an important role in such multi-modal survival modeling. 
The univariate Cox model \eqref{eq:Cox} can be straightforwardly generalized by including multiple clinical, demographic or omics variables of interest, as long as the total number of covariates is much smaller than the number of samples; this often requires the use of variable or feature selection methods.
\cite{Heinze2018} provides pragmatic recommendations for practitioners of multivariable analysis with variable selection methods in low-dimensional modeling problems.

\section{Survival analysis with high-dimensional input data}\label{sec:multivariate}

Since some omics data contain high-number of variables (e.g., RNA-seq with ca. 60000 transcriptomic features and DNA methylation with ca. 450000 features), there is a need to reduce the computational and modeling burden in multivariable analyses. 
One heuristics approach to pre-select a subset of features is to include only omics features at a specific statistical significance level when fitting a univariate Cox model \eqref{eq:Cox}; the pre-specified significance level would usually be higher than the commonly used 0.05 threshold, e.g., 0.1 or 0.2, to avoid losing important features. 
However, this univariate approach focuses on features that are independently associated with the outcome and might miss variables that are important in combination with other features \citep{Okser2013}. 

Another simple approach is to pre-select omics features by variance, since larger variability across patients usually implies higher biological information, or at least predictive signal. 
One can for example pre-select omics features explaining 50\% of the total cumulative variance \citep{Zhao2020}. 
Such unsupervised feature pre-selection is a recommended method to reduce dimensionality when dealing with hundreds or thousands of omics features, with the aim to improve the stability of final feature selection. 
For example, \cite{Bommert2022} showed that the simple variance filter was the best method among all considered filter methods in terms of the predictive accuracy, run time and the feature selection stability in their benchmark study. 
\cite{Zhao2023} showed that the stability of final feature subset critically depends on the pre-selected feature set when using a standard Bayesian stochastic search variable selection method (see Section ``\hyperref[sec:bayes]{Supervised learning via Bayesian priors}"), and their proposed method that used known biological relationships between omics features lead to a more stable feature selection and slightly improved outcome prediction.

When drawing conclusions on survival differences solely from the univariate Cox model \eqref{eq:Cox}, it is important to adjust the $p$-values of risk features for multiple comparisons to control false positives globally, e.g., by controlling the family-wise error rate (FWER) or false discovery rate (FDR). 
The problem of multiple testing is beyond the scope of this tutorial, but the interested readers are referred to a recent review article \cite{Korthauer2019}. 
We note also that univariate analysis does not consider any relationships between multiple omics features, e.g., potential confounders for survival outcomes \citep{Clark2003}.
To avoid making seriously misleading conclusions in such cases, it is necessary to perform multivariable survival analysis \citep{Bradburn2003}.

Omics data can have hundreds of thousands of variables measured at various molecular levels, which greatly challenges the classical multivariable regression models for time-to-event endpoints, since the number of variables is often much larger than the number of patients (i.e., $p\gg n$). 
To proceed with survival analysis, one option is to reduce the dimensionality of the omics features via unsupervised learning, and then investigate the association of the learned low-dimensional variables with survival outcomes (see next section).
An alternative approach is to directly use supervised learning methods, such as penalized regressions or sparse Bayesian models (\hyperref[tab:superviseCox]{Table~\ref*{tab:superviseCox}}), which enable the modeling of high-dimensional omics features, and the selection of key important features associated with the survival outcomes.

\subsection{Unsupervised learning}

Unsupervised learning methods aim to identify hidden patterns or data groupings, and are for example useful when a phenotype (e.g., a cancer type) is to be divided into several subtypes (e.g., to explain heterogeneity among patients). 
For example, breast cancer has been traditionally categorized into five conceptual molecular classes, originally using pairwise average-linkage cluster analysis of DNA microarray data, to better understand tumor biology and guide clinical decision making \citep{Perou2000}. 
Unsupervised methods learn underlying patterns from unlabeled data by transforming high-dimensional omics features into a lower dimensional space. 
Principal component analysis (PCA) is a classical multivariate technique that represents high-dimensional features in a low-dimensional space by building orthogonal (uncorrelated) linear combinations of the features that best capture the variance in the high-dimensional data. 

Different from the distance-based PCA with linear transformation, non-linear techniques have recently emerged, such as $t$-stochastic neighbour embedding ($t$-SNE). 
$t$-SNE uses pairwise similarities of individuals based on Kullback-Leibler divergence to project similarities into a lower dimensional space, ensuring that individuals with similar omics features are close in the generated embedding \citep{vandermaaten2008}. 
The focus of $t$-SNE is on preserving local distances between neighbouring data points.
Another non-linear alternative to PCA is UMAP (Uniform Manifold Approximation and Projection), which is a general dimension reduction method built upon Riemannian geometry \citep{McInnes2018}. 
Compared to other non-linear methods of dimension reduction such as $t$-SNE, UMAP can sometimes provide better visualization quality in a shorter amount of time, while preserving the global structure of the omics data better. 

Unsupervised methods only make use of the input data matrix $\mathbf{X}$ and are agnostic to the survival information. 
To find out whether a given omics profile (i.e., the full set of omics features, not individual features) is associated with survival outcomes, a straightforward approach is to use a few representative components from PCA, $t$-SNE or UMAP as covariates in a multivariable Cox model. 
An alternative is to use semi-supervised methods \citep{Bair2004} that combine the clustering procedure and survival modeling together. 
However, such principal component regression and semi-supervised methods lose interpretability of the individual omics features, since each component is a linear or nonlinear combination of all omics features.

\begin{table*}[t]
  \caption{\it \centering{Cox-type supervised learning methods}\label{tab:superviseCox}}
  {\tiny\makebox[\textwidth][c]{
  \begin{tabular}{l c c c c}  
  \toprule
  {\bf \ \ \ \ Method} & {\bf Feature selection via} & {\bf Grouping} & {\bf Uncertainty} & {\bf Comment} \\
  & & {\bf effects} & {\bf quantification} \\
  \midrule
  {\bf Penalized regressions:} & {\bf penalty}
  \smallskip\\
  \ \ Lasso Cox [\cite{Tibshirani1997}]       & $\ell_1$-norm   & \xmark  & \xmark &
  \vspace{2pt}\\
  \ \ Adaptive Lasso Cox [\cite{Zhang2007}]       & weighted $\ell_1$-norm   & \xmark  & \xmark & less false positives than Lasso
  \vspace{2pt}\\
  \ \ Elastic Net Cox [\cite{Simon2011}]     & $\ell_1/\ell_2$-norm   & \cmark  & \xmark &
  \vspace{2pt}\\
  \ \ Group-Lasso Cox [\cite{Kim2012}]     & $\ell_2$-norm   & \cmark  & \xmark & independent groups of features selected
  \vspace{2pt}\\
  \ \ Sparse Group-Lasso Cox [\cite{Simon2013}]     & $\ell_1/\ell_2$-norm   & \cmark  & \xmark &
  \vspace{2pt}\\
  \ \ SCAD Cox [\cite{Fan2002}]      & quadratic spline and symmetric penalty   & \cmark  & \xmark &  selection of relatively large effects
  \vspace{2pt}\\
  \ \ SIS Cox [\cite{Fan2010}]    & top ranked variables and any penalty & \cmark  & \xmark & suited to ultra-high dimensions
  \smallskip\\
  {\bf Bayesian models:} & {\bf shrinkage prior}
  \smallskip\\
  \ \ [\cite{Lee2011}]     & Lasso (Laplace) prior   & \xmark  & \cmark &  selection of posterior mean with a cutoff
  \vspace{2pt}\\
  \ \ [\cite{Lee2015}]     & Elastic Net, group/fused Lasso priors   & \cmark  & \cmark & selection of posterior mean with a cutoff
  \vspace{2pt}\\
  \ \ [\cite{Konrath2013}]     & Spike-and-slab prior   & \xmark  & \cmark &
  \vspace{2pt}\\
  \ \ [\cite{Madjar2021}]     & Spike-and-slab \& MRF priors   & \cmark  & \cmark &
  \vspace{2pt}\\
  \ \ [\cite{Mu2021}]     & Horseshoe prior   & \xmark  & \cmark & selection of posterior mean with a cutoff \\
  \botrule
  \end{tabular}
  }}
\end{table*}

\subsection{Supervised learning via penalized regressions}\label{sec:penalized}

\begin{figure}[!h]
  \centering
  \includegraphics[height=0.28 \textwidth]{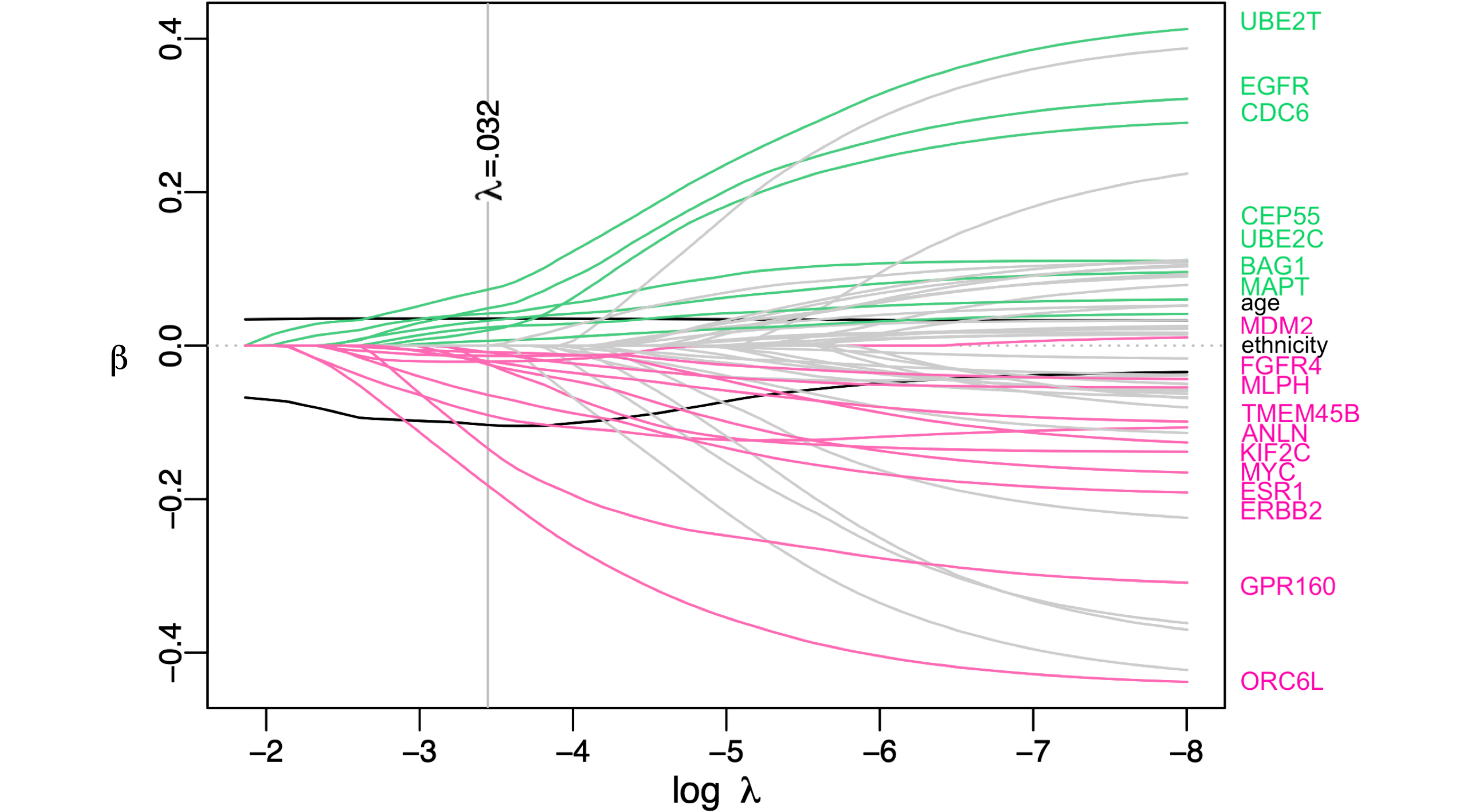}
  \caption{\it Coefficient trace plot of an Elastic Net Cox model for overall survival prognosis of breast cancer patients from TCGA based on transcriptomic data and mandatory demographic variables age and ethnicity. The y-axis shows the magnitude of each feature's coefficient given the strength of penalization displayed on the x-axis (from left to right the penalization decreases). The vertical gray line indicates the optimal $\lambda=0.032$ (maximizes the partial likelihood via cross-validation) and its corresponding selected features are marked with green (positive coefficient) and red (negative coefficient) colors. Note that the demographic variables age and ethnicity were not penalized, so that their coefficient paths (black color) did not start from zero in the figure.} \label{fig:elastic}
\end{figure}

For the purpose of personalized cancer medicine, one is typically interested in identifying risk factor combinations from clinical and omics features.
These factors can be targeted (directly or indirectly) via therapeutic strategies or used for diagnostics. 
Therefore, the objective is to identify a parsimonious set of features linked to survival outcomes by utilizing the wealth of information present in, for example, the vast amount of available omics data. 
Penalized Lasso Cox regression \citep{Tibshirani1997} can be used to select a few relevant omics features by estimating their coefficients as non-zero (the non-relevant features' coefficients are shrunk to zero) via maximizing the penalized partial log-likelihood function of the regression coefficients with $\ell_1$-norm penalty
\begin{equation*}
2/n \cdot \ell(\bm\beta|\mathcal D) - \lambda \|\bm\beta\|_1 \label{eq:Lasso}\tag{2}
\end{equation*}
Here, $2/n$ is a scaling factor for convenience, $\mathcal D=\{(\tilde T_i, \delta_i, \mathbf X_i): i=1,\cdots,n\}$, $\mathbf{X}_i$ includes $p$ (omics) features of the $i$-th patient, $\lambda$ is a tuning parameter to control the overall penalty strength of the coefficients, $\|\bm\beta\|_1 = \sum_{j=1}^p |\beta_j|$, and the partial log-likelihood is
$\ell (\bm\beta|\mathcal D)  = \log \prod_{i=1}^n \left\{\frac{\exp(\mathbf{X}_i\bm\beta)}{\sum_{l\in \mathcal{R}_k }\exp(\mathbf{X}_l\bm\beta)}\right\}^{\delta_i},$ 
where $\mathcal{R}_k = \{l : Y_l(T_k) =1\}$ is the risk set at time $T_k$. 
The Elastic Net Cox model \citep{Simon2011} considers both the Lasso feature selection and the grouping effect of correlated omics features in ridge regression via a combination of the $\ell_1$- and $\ell_2$-norm (i.e., $\ell_1/\ell_2$-norm) penalty $\lambda \left\{\alpha\|\bm\beta\|_1 + \frac{1}{2}(1-\alpha)\|\bm\beta\|_2^2 \right\}$ (where $\|\bm\beta\|_2^2 = \sum_{j=1}^p |\beta_j|^2$) , which can usually improve the prediction performance over the Lasso Cox model. 
\hyperref[fig:elastic]{Figure~\ref*{fig:elastic}} shows an example of Elastic Net Cox model feature selection from gene expression features associated with breast cancer patients' survival. 
Note that often we wish to include a small set of well-established clinical risk factors in a survival model. 
Since they are established as important covariates, they can be included in the Lasso or Elastic Net Cox model as mandatory covariates without penalization. 
Then the penalized partial log-likelihood function becomes
\begin{equation*}
\frac{2}{n}\log \prod_{i=1}^n \left\{\frac{\exp(\mathbf{X}_{0i}\bm\beta_0 + \mathbf{X}_i\bm\beta)}{\sum_{l\in \mathcal{R}_k }\exp(\mathbf{X}_{0l}\bm\beta_0 + \mathbf{X}_l\bm\beta)}\right\}^{\delta_i} - \text{pen} (\bm\beta), \label{eq:LassoB0}\tag{3}
\end{equation*}
where $\bm\beta_0$ are coefficients corresponding to the $i$-th individual's mandatory covariates $\mathbf{X}_{0i}$, pen($\bm\beta$) is a $\ell_1$- or $\ell_1/\ell_2$-norm penalty for feature selection of omics features $\mathbf{X}_i$. 
\cite{DeBin2014} investigated more strategies to combine a low-dimensional set of well-established clinical factors and high-dimensional omics features into a global prediction model.

There are many alternative penalties that will achieve feature selection which have been applied to Cox proportional hazards regression for survival outcomes, such as the Adaptive Lasso Cox model that incorporates different penalties for different coefficients to retain important variables \citep{Zhang2007}, 
Group-Lasso that performs group selection on (predefined) groups of variables \citep{Kim2012},
Sparse Group-Lasso that introduces sparse effects both on a group and within group level \citep{Simon2013}, the smoothly clipped absolute deviation (SCAD) Cox model that overcomes substantially biased estimates for large coefficients in ultra-sparse models \citep{Fan2002}, 
sure independence screening (SIS) procedure in combination with Cox model that speeds-up the feature selection dramatically and can also improve the accuracy of estimation when dimensionality becomes ultra-high, i.e., $\log(p)=\mathcal O(n^\xi)$ for some $\xi>0$ \citep{Fan2010}. 
However, all these penalized Cox models do not directly provide uncertainty of feature selection or survival prediction. 
One empirical way for uncertainty quantification is through additional resampling-based methods, see Section ``\hyperref[sec:validate]{Survival model validation}'' for more details.

\subsection{Supervised learning via Bayesian priors} \label{sec:bayes}

\begin{figure}[!h]
  \centering
  \includegraphics[height=0.13 \textwidth]{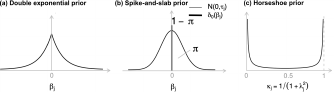}
  \caption{\it Point estimation and uncertainty quantification of regression coeﬀicients using Bayesian Lasso Cox model with Laplace prior for overall survival prognosis of breast cancer patients from TCGA based on transcriptomic data and mandatory demographic variables age and ethnicity. Solid dots indicate the posterior mean over 20000 MCMC iterations excluding burn-in period, and horizontal lines show the corresponding 95\% credible intervals.} \label{fig:bayesLasso}
\end{figure}

Bayesian inference is an appealing approach for survival analysis due to its ability to provide straightforward uncertainty quantification (e.g., credible intervals) of parameters and survival probabilities. 
For instance, \cite{Lee2011} proposed a Bayesian version of the Lasso Cox (Bayesian Lasso Cox) model that provides credible intervals of coefficients fairly straightforward (see \hyperref[fig:bayesLasso]{Figure~\ref*{fig:bayesLasso}}), but it is not easy to derive confidence intervals of coefficients in a Lasso-type model. 

The fundamental theorem of Bayesian methods is Bayes' rule. Let $\bm\beta$ be the parameters of interest, e.g., gene effects, external to the data $\mathcal D$. 
To estimate the parameters $\bm\beta$ given the data information, one can use Bayes' rule to obtain the conditional (posterior) distribution of $\bm\beta$:
$$f(\bm\beta | \mathcal D)  
= \frac{f(\mathcal D |\bm\beta)f(\bm\beta)}{f(\mathcal D)} 
\propto f(\mathcal D |\bm\beta)f(\bm\beta),$$
where $f(\mathcal D)$ is a normalization constant that can be neglected in inference since the data are already observed, $f(\mathcal D |\bm\beta)$ is the likelihood of the data viewed as a function of the parameters of a statistical model and $f(\bm\beta)$ is the prior distribution of $\bm\beta$. 
The prior distribution can be chosen either based on historical data from past similar studies or from popular (non-)informative priors \citep{Ibrahim2001}. 
The estimation of $\bm\beta$ is to maximize the posterior $f(\bm\beta | \mathcal D)$ or $\log f(\bm\beta | \mathcal D)$, which is equivalent to maximize the sum of the log-likelihood and the log prior, i.e., $\log f(\bm\beta | \mathcal D) = \log f(\mathcal D |\bm\beta)+\log f(\bm\beta)$, which takes into account both the observed data information and the prior information in an optimal way. 

\begin{figure}[!h]
  \centering
  \includegraphics[height=0.4 \textwidth]{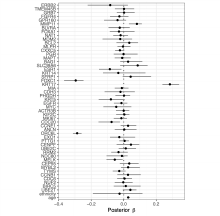}
  \caption{\it Density of shrinkage priors in Bayesian survival modeling. (a) Density of the double exponential (Laplace) prior. (b) Density of the mixture spike-and-slab prior. The spike component $\delta_0(\beta_j)$ induces $\beta_j=0$ and the slab component $\mathcal N(0,\tau_j)$ induces $\beta_j\ne 0$. (c) Density of the horseshoe prior. The shrinkage weight $\kappa_j$ close to $0$ shrinks $\beta_j$ towards zero, and $\kappa_j$ close to $1$ allows $\beta_j$ to escape the shrinkage.} \label{fig:spike-slab}
\end{figure}

The Bayesian version of the Lasso Cox model can have a log posterior similar to the frequentist penalized partial log-likelihood function (\ref{eq:Lasso}), if we assign independent double exponential (also known as Laplace, \hyperref[fig:spike-slab]{Figure~\ref*{fig:spike-slab}a}) prior, $f(\bm\beta)=\prod_{j=1}^p\frac{\lambda}{2}\exp\{-\lambda|\beta_j|\}$ with a scale parameter $\lambda>0$, for all the coefficients, i.e., 
\begin{equation*}
\log f(\bm\beta | \mathcal D, \lambda) = \ell^*(\bm\beta | \mathcal D) - \lambda\|\bm\beta\|_1 + C,\label{eq:bayesLasso}\tag{4}
\end{equation*}
where $\ell^*(\bm\beta | \mathcal D)$ is the full log-likelihood function i.e., $\log(\int_0^t h_0(s)ds)+\ell(\bm\beta | \mathcal D)$, $C$ is a normalization constant independent of $\bm\beta$ and the $\ell_1$-norm penalty tends to choose only a few nonzero coefficients. 
Markov chain Monte Carlo (MCMC) sampling can be performed for posterior inference of $\bm\beta$. 
Note that instead of using the partial log-likelihood $\ell(\bm\beta | \mathcal D)$ in \eqref{eq:bayesLasso}, a full log-likelihood function is used, which includes the baseline hazard function. 
This can be achieved, for example, by assigning a stochastic process prior, e.g., a gamma process, for the cumulative baseline hazard function. 
More details about the prior setup and inference can be found in \cite{Lee2011}. 
\cite{Lee2015} extended the Laplace prior to Elastic Net prior, fused Lasso prior and group Lasso prior, which are often more suitable for correlated omics features in survival analysis. 
But \cite{Lee2011} and \cite{Lee2015} assign the same shrinkage priors to all covariates indiscriminately, see \cite{Zucknick2015} for an extended Bayesian Lasso Cox model which permits the use of mandatory covariates. 

Note that the Bayesian version of the Lasso with Laplace-type priors in practice do not result in automatic feature selection, because only the posterior modes of the coefficients are equivalent to the frequentist Lasso solution, while in Bayesian inference one usually focuses on the posterior means as point estimates. 
As an alternative, a particular omics feature can be excluded if the estimated credible interval of the corresponding coefficient covers zero. 

Stochastic search variable selection (SSVS) is an alternative approach to identify important covariates \citep{George1993,Konrath2013}.
SSVS uses independent spike-and-slab priors for regression coefficients, e.g.
\begin{equation*}
\beta_j | \gamma_j, \tau_j^2 \sim \gamma_j\mathcal N(0, \tau_j^2) + (1-\gamma_j)\delta_0(\beta_j),
\label{eq:spike-slab}\tag{5}
\end{equation*}
where $\gamma_j$ ($j=1,\cdots,p$) is a latent variable (which can have a Bernoulli prior with a fixed probability $\pi$) for feature selection indicating $\beta_{j}\neq 0$ if $\gamma_{j}=1$ and $\beta_{j}= 0$ if $\gamma_{j}=0$, $\tau_j^2$ is an additional shrinkage parameter which can be assigned with an additional prior (e.g., exponential or inverse gamma prior), and $\delta_0(\cdot)$ is the Dirac delta function. 
\hyperref[fig:spike-slab]{Figure~\ref*{fig:spike-slab}b} shows the two components of the spike-and-slab mixture distribution. 
Recently, \cite{Madjar2021} proposed graph-structured feature selection priors for Cox model by assigning a Markov random field prior on the latent variables, in which the graph helps to identify pathways of functionally related genes or proteins that are simultaneously prognostic in different patient cohorts. 
Formulation (\ref{eq:spike-slab}) implies independence between the priors of the individual $\beta_j$ in the slab component. In contrast, a variant of the spike-and-slab prior has a $g$-prior slab \citep{Zellner1986,Held2016} 
$$\bm\beta_{\bm\gamma} |\bm\gamma \sim \mathcal N(\bm 0, g\mathcal{I}_{\bm\beta_{\bm\gamma},\bm\beta_{\bm\gamma}}^{-1}),$$ 
where $\bm\beta_{\bm\gamma} = \{\beta_j:\gamma_j=1,j=1,\cdots,p\}$, $\bm\gamma = \{\gamma_j:j=1,\cdots,p\}$, $g$ is either a scalar estimated by Empirical Bayes or assigned with additional prior, and $\mathcal{I}_{\bm\beta_{\bm\gamma},\bm\beta_{\bm\gamma}}^{-1}$ is the expected Fisher information for $\bm\beta_{\bm\gamma}$.

Another popular shrinkage prior is the horseshoe prior, a continuous and global-local shrinkage prior, in which the global parameters allow for sharing information across omics features and the local parameters allow for adjustments at the individual omics feature level \citep{Carvalho2009,Mu2021}. 
In a similar setup to the Cox model with spike-and-slab priors in \eqref{eq:spike-slab}, a horseshoe prior for the regression coefficient is 
\begin{equation*}
\beta_j|\lambda^2,\tau_j^2 \sim \mathcal N(0, \lambda_j^2\tau^2 ),\ \ \lambda_j \sim \text{C}^+(0,1),\ \ \tau \sim \text{C}^+(0,1),
\end{equation*}
where the local parameter $\lambda_j$ and global parameter $\tau$ are both half-Cauchy distributed $\text{C}^+(\cdot,\cdot)$. 
With the horseshoe prior, the posterior mean of $\beta_j$ will be shrunk by a weight $\kappa_j=\frac{1}{1+\lambda_j^2}\in (0,1)$ as in \hyperref[fig:spike-slab]{Figure~\ref*{fig:spike-slab}c}, where $\kappa_j\to 0$ induces $\beta_j\to 0$. 
Using a user-adjustable cutoff value, many coefficients can be shrunk to zero, enabling the selection of only a few associated omics features (with non-zero coefficients). 

Although Bayesian models can quantify uncertainty of estimators more straightforward than penalized regressions, most Bayesian Cox-type models for high-dimensional covariates do not have user-friendly and standalone R packages on CRAN or GitHub. 
The main reason is the high computational cost of running a high-dimensional Bayesian Cox model. 
Advanced users with programming capabilities can contact the corresponding authors for original scripts. 
Since Bayesian priors are more flexible than frequentist Lasso-type penalties, it can be easier to extend Bayesian models by changing shrinkage priors while keeping almost the same algorithm framework. 
This means that the Bayesian framework can be more suitable if one is interested in tailoring the shrinkage effects, e.g., to include prior knowledge about the importance of omics features, for example features corresponding to a molecular pathway that is known to be affected in the disease under study. 
For more information on different Bayesian priors in cancer prognosis, we suggest a recent review by \cite{Chu2022} which summarized other different shrinkage priors on regression coefficients, such as Gaussian-gamma, Gaussian, Cauchy, pMOM (product moment distribution), piMOM (product inverse moment distribution) and peNMIG (parameter-expanded normal-mixture-of-inverse-gamma distribution) priors.

\section{Survival model validation}\label{sec:validate}

Model validation plays an important role in identifying potential issues, such as model misspecification or overfitting. 
This is achieved by revisiting the model's specifications and assumptions following model estimation. 
For example, the Cox model \eqref{eq:Cox} requires proportional hazards and the logarithm of the hazard to be linear with respect to the model covariates. 
The former assumption can for example be checked by the cumulative Schoenfeld residuals \citep{Grambsch1994}, and the latter assumption by plotting a nonlinear functional form (e.g., spline) for the effect of a covariate. 
If the Cox model assumptions are not satisfied, one can try certain transformations of covariates (e.g., Box-Cox power transformations, \citealp{Box1964}), allow time-varying coefficients or model interactions among covariates \citep{Ng2023},
or investigate patient stratification using unsupervised approaches \citep{Cristescu2015}. 
However, the assumption checks are usually suitable only for low-dimensional models, i.e., for a few clinical variables or a few factors projected from the high-dimensional omics feature space. 
Novel approaches for assumption checks in general high-dimensional settings require further methodological developments. 
\cite{Johnson2011} used heuristic methods to investigate the Cox model assumptions by separately fitting univariate Cox models one feature at one time, and check $p$-values for the score tests of individual features and $p$-values for testing the proportional hazards assumption of univariate Cox models. 
But the univariate Cox models do not take into account confounding variables. 
An alternative approach is to loosen the model assumptions for a more robust modeling approach. 
One example developed specifically for feature selection under possibly non-proportional hazards in a high-dimensional space is concordance regression \citep{Dunkler2010}. 

\subsection{\bf Feature stability analysis}\label{sec:stability}

One important aspect in model validation when using omics or other high-dimensional data is the potential instability of feature selection \citep{Kalousis2007}. 
Feature selection using penalized regressions as described in Section ``\hyperref[sec:penalized]{Supervised learning via penalized regressions}" heavily depends on the values of the penalty parameters (e.g., for the $\lambda$ parameter in Lasso Cox model \eqref{eq:Lasso}). 
The penalty parameters are often optimized by cross-validation (CV) or other resampling methods, and the uncertainty associated with the random selection of subsets may result in uncertainty in the feature selection, e.g., different CV folds will typically result in different selected features. 
A straightforward way to identify the most stable features is to find the overlap of identified omics features among different data subsets (e.g., CV folds or resamples) to avoid high false discovery rate \citep{Zucknick2008}.
One can also perform stability selection \citep{Meinshausen2010}, which allows to select the most stable features at a given Type-I error level for a Lasso or Elastic Net Cox model \citep{Sill2014}.

For the Bayesian models in Section ``\hyperref[sec:bayes]{Supervised learning via Bayesian priors}", feature selection stability is naturally assessed by the uncertainty of coefficients' estimators, as reflected in the posterior variances of $\beta_j$ or the posterior selection probabilities $p(\gamma_j | \mathcal D)$ (in SSVS), which is a natural benefit of utilizing full Bayesian inference. 
Although the uncertainty in feature selection introduces increased variability in the predicted survival probabilities, in the Bayesian framework, this can be addressed quite naturally by averaging the survival predictions over all models using Bayesian model averaging \citep{Volinsky1997}. 
If one is interested in a single model, rather than model averaging, the median probability model \citep{Barbieri2004} can be used for uncertainty analyses in survival and high-dimensional omics data \citep{Madjar2021}.

\subsection{\bf Survival prediction and calibration}

The fundamental goal of any statistical prediction model is to achieve a better prediction performance than an existing statistical model which we could call the ``conventional model'' \citep{Gerds2021}. The special nature of the combination of clinical and/or other known prognostic factors (typically low-dimensional, with established effect) and novel omics features (high-dimensional, with unknown effect) should especially be taken into account. Therefore, a survival model consisting of only established clinical and/or other known prognostic factors should serve as the benchmark (i.e., conventional model) for the upcoming modeling. The inclusion of new covariates (e.g., omics features) into a prognostic model only makes sense if the new covariates added prognostic value over the established clinical prognostic factors \citep{DeBin2014}, i.e., the new prognostic model consisting of the new covariates plus benchmark covariates improves the prediction performance over the conventional model. 

To confirm that the identified clinical and omics features have prognostic power with respect to the prediction of patients' survival outcomes, a model should be both accurate (low prediction error) and precise (low prediction uncertainty). 
The simplest way to demonstrate the prognostic power is to dichotomize the prognostic scores (i.e., linear predictor $\mathbf X_i\bm\beta$ in the Cox model, \cite{Cox1957}) by its median value, and then use a log-rank test to compare the survival probabilities of the patients in the two groups, see \hyperref[fig:KM]{Figure~\ref*{fig:KM}b}. 
Similarly, one can categorize the prognostic scores by multiple quantiles (e.g., 25\%, 50\% median, and 75\%) into multiple groups of patients and perform a log-rank test. 

To validate a prediction model systematically \citep{Rahman2017,Royston2013}, the predictive performance of the model is commonly addressed by 
\begin{itemize}
  \item {\bf discrimination}: the ability of the model to distinguish between low and high risk patients,
  \item {\bf calibration}: the agreement between the observed outcomes and predicted survival probabilities, and
  \item {\bf overall performance}: the distance between the observed and predicted survival probabilities.
\end{itemize}

\subsubsection{Discrimination performance}

If one focuses on survival prediction at a fixed time point (e.g., 5-year survival probability), a receiver operating characteristic (ROC) curve can be used to evaluate the prognostic (i.e., prediction or discrimination of survival) ability of the survival model, often summarized by its area under the ROC curve (AUC) \citep{Heagerty2000}, see \hyperref[fig:calibration]{Figure~\ref*{fig:calibration}a} for an example. 
An AUC of 0.5 is equivalent to the toss of a coin, and the closer the AUC is to 1, the more predictive is the model. 
When making predictions at multiple time points, ROC curves can be summarized as time-dependent AUC scores, i.e., AUC scores calculated at prespecified time points. 
Alternatively, the concordance index (C-index) provides a more global, time-independent assessment of the discrimination ability of a prognostic model, such that a better model predicts higher prognostic scores for patients with shorter survival times \citep{Harrell1982, Antolini2005}, i.e., 
\begin{equation*}
C = \mathbb P\{S(t| \mathbf X_i(\mathbf t)) < S(t| \mathbf X_j(\mathbf t))\ | T_i <T_j\ \& \ \delta_i =0\},
\end{equation*}
which means in the absence of censoring, any pair of individuals $\{i,j\}$ with survival times $T_i<T_j$ is concordant if and only if $S(t| \mathbf X_i(\mathbf t)) < S(t| \mathbf X_j(\mathbf t))$ for any $t$ (equivalent to ranking the prognostic scores $\mathbf X_i\bm\beta > \mathbf X_j\bm\beta$ in a Cox model), where $\mathbf t$ denotes the time instants where there are covariate variations. 
The C-index can be expressed as a weighted average of the time-dependent AUC over time \citep{Heagerty2005}. 
Therefore, its interpretation is similar to the AUC, where a C-index of 0.5 indicates random predictions, while a perfect prognostic model would have a C-index of 1. 
There are multiple types of C-indices for survival modeling, in particular the most frequently used Harrell's \citep{Harrell1982} and Uno's C-index \citep{Uno2011}. 
Uno's C-index is more robust than Harrell's C-index, in case there is dependence of the concordance on the study-specific censoring distribution. 

In the classical Cox modeling framework, both Harrell's and Uno's C-indices only depend on the linear predictors $\mathbf X_i\bm\beta$, which is independent of $t$. 
But if a model includes covariates with time-dependent effects $\bm\beta(t)$ and/or time-dependent covariates $\mathbf X_i(t)$, Harrell's and Uno's C-indices are difficult to be calculated, since they require the calculation of survival functions for each individual over time. 
In this context, \cite{Antolini2005} proposed a time-dependent C-index, which assesses the concordance of a model's survival distribution predictions $\hat{S}(t)$. 
This means that Antolini's C-index requires the full specification of $\hat{S}(t)$, even though a C-index only compares the survival probabilities between any pair of individuals, that is, it only assesses whether the relative order of estimated survival probabilities is concordant with observed survival outcomes \citep{Blanche2019}. 
Time-dependent prediction indices can better evaluate a model including candidate features with time-dependent effects and/or time-dependent features. 
To avoid C-hacking among different C-indices in model comparison, \cite{Sonabend2022} recommended that if all models make survival distribution predictions, then select a time-dependent C-index; otherwise choose a time-independent measure (e.g., Uno's C-index); if there is a combination of risk- and distribution-predicting models, then choose a transformation method for analysis (e.g., expected mortality).

\subsubsection{Calibration performance}

Calibration is to quantify the agreement between the observed and predicted outcomes, which is useful for both internal and external model validation and is recommended to report routinely. 
The calibration slope is commonly used \citep{vanHouwelingen2000}, which is the slope of the regression of the observed/actual survival probabilities on the model-predicted survival probabilities.
A survival model can be reported with the estimated $t$-year survival probabilities in predefined subgroups, denoted as $S_{\text{model}}(t|g)$ for subgroups $g=1,\cdots,G$. 
The observed $t$-year survival probabilities in the subgroups can be estimated by the KM method, denoted as $S_{\text{KM}}(t|g)$. 
Using the $\ln(-\ln(\cdot))$-link, the calibration model is
\begin{equation*}
\ln(-\ln(S_{\text{KM}}(t|g))) = \alpha + \beta \ln(-\ln(S_{\text{model}}(t|g))) + \epsilon, \label{eq:calibration}\tag{6} 
\end{equation*}
where $\epsilon$ is an error term. If the intercept $\alpha=0$ and the slope $\beta=1$, it means that the survival prediction model is well calibrated. 
For example, \hyperref[fig:calibration]{Figure~\ref*{fig:calibration}b} shows a calibration plot, visualizing the calibration of the estimated 5-year survival probabilities (with 95\% confidence interval by bootstrapping) using the KM method for TCGA-BRCA patients grouped by the quartiles of Cox-model predicted survival probabilities. 
Furthermore, one can calibrate a Cox model in terms of the baseline cumulative hazard and prognostic score. 
For non-proportional hazard models, calibration using the model cumulative hazard function can be considered \citep{vanHouwelingen2000}. 

As an alternative to the calibration slope at a single time point, \cite{Andres2018} and \cite{Haider2020} suggested the distributional (D)-calibration for accounting survival probabilities across all time points. This can be useful when assessing the entire post-treatment survival prediction, for example, assessing post liver transplantation survival utility in \cite{Andres2018}.

\subsubsection{Overall performance}

Scoring rules can evaluate the accuracy and confidence of probabilistic predictions, and assess both discrimination and calibration \citep{Gneiting2007,Avati2020}. 
The idea of scoring rules dates back to \cite{Brier1950} which assigned a numerical score for verifying ensemble-based probabilistic predictions of discrete outcomes. 

\cite{Graf1999} proposed the time-dependent Brier score, which is the expected mean-squared error of survival probability prediction at different time points, i.e.
{\medmuskip=-2mu\nulldelimiterspace=-1pt\scriptspace=0pt
\begin{equation*}
\text{BS}(t) = \frac{1}{n} \sum_{i=1}^n \left\{ \frac{\hat S(t|t_i)^2 \mathbbm{1}\{t_i \le t\ \&\ \delta_i=1\}}{\hat G(t_i)}\ +\ \frac{\{1-\hat S(t|t_i)\}^2 \mathbbm{1}\{t_i > t\}}{\hat G(t)} \right\},
\end{equation*} 
}\noindent
where $t_i$ is the survival time of $i$-th individual, $\hat S(t|t_i)$ is the Cox-model predicted survival probability and $\hat G(t)$ is the KM estimate of the censoring distribution. 
The benefit of the Brier score is that it does not only measure discrimination, similar to evaluation measures like the C-index, but also calibration performance of a survival model. 
The integrated Brier score (IBS) is as a single measure of prediction accuracy integrating BS($t$) over an entire follow-up time period. 
\cite{Hielscher2010} presented a comparison between the IBS and a $D$ measure \citep{Schemper2000}, which is an integrated measure based on the mean absolute deviation rather than the mean-squared error used in IBS. 
The $D$ measure is more robust towards extreme observations and has a smaller variance than the IBS. 

To overcome potential overfitting when using feature selection and model estimation, the survival predictions and model calibration should be evaluated in independent validation data sets. 
As independent validation data are seldom available, we can split the available data into training and validation data sets. However, any single split of the data hides partial data from all steps of model building, which might introduce bias, it is recommended to use resampling-based methods for assessing the survival model’s performance. In addition, using resampling-based methods also allow us to estimate the uncertainty of the performance estimator (\citealp{Sill2014,Gerds2021}, chapter 7). 
This can be done for example by repeatedly splitting the data to training and validation sets, and evaluating a survival model’s performance on the different validation sets using various discrimination or calibration indices. 
The .632+ bootstrap estimator for a discrimination or calibration index can balance the apparent (training) error and the average out-of-bag bootstrap error, and in addition accounts for the relative overfitting based on a no-information error rate in high-dimensional settings \citep{Schumacher2007,Binderbootstrap2008}. 
This is a typical machine learning approach with two levels of resampling. 
The outer layer of resampling is to evaluate the prediction performance and the inner layer of resampling (usually CV) is to optimize model's tuning parameters. 
We note that the preparatory steps, such multi-modal data standardization and feature pre-selection in the context of high-dimensional input data, may affect the survival prediction performance, and thus should be included in the resampling steps for model validation.

\subsubsection{Graphical representation}

\begin{figure}[!h]
  \centering
  \includegraphics[height=0.3 \textwidth]{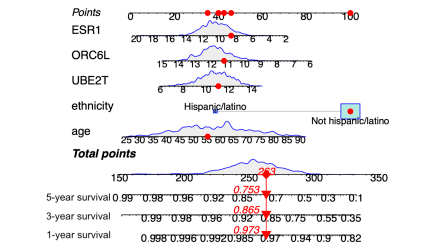}
  \caption{\it Nomogram developed to estimate the overall survival probability for TCGA-BRCA patients based on clinical (age) demographic (ethnicity) and three selected mRNA features form a Lasso Cox model. The red coloured symbols represent example patient’s information and predicted probabilities of 1-year, 3-year and 5-year survival.}\label{fig:nomogram}
\end{figure}

After confirming that a model is valid (assumptions hold), accurate (low prediction error), precise (uncertainty of performance measures properly quantified) and its predictions generalizable beyond the training data set (using independent validation data if available), a prognostic nomogram \citep{Kattan1999} can be used to summarize the prognostic effect of the identified clinical and omics features on the risk of a specific year's overall survival (\hyperref[fig:nomogram]{Figure~\ref*{fig:nomogram}}), which may help the clinicians to enhance the patient management and personalized treatment strategies. 
For example, the red colored dots in \hyperref[fig:nomogram]{Figure~\ref*{fig:nomogram}} show the information of the identified five variables from an example patient and the corresponding scoring points. 
The summed scoring points of 263 maps to the predicted 1-year, 3-year and 5-year survival probabilities of this patient. 
Note that most nomograms treat the identified variables independently in the risk calculation, even though there may be significant interactions among the model features that were used in the feature selection step. 
However, visualizing such interaction effects would make the nomograms less accessible and interpretable, and so, there is still a room for improvement in how to translate the multivariate risk scores into clinical practice. 

\begin{figure}[!h]
  \centering
  \includegraphics[height=0.25 \textwidth]{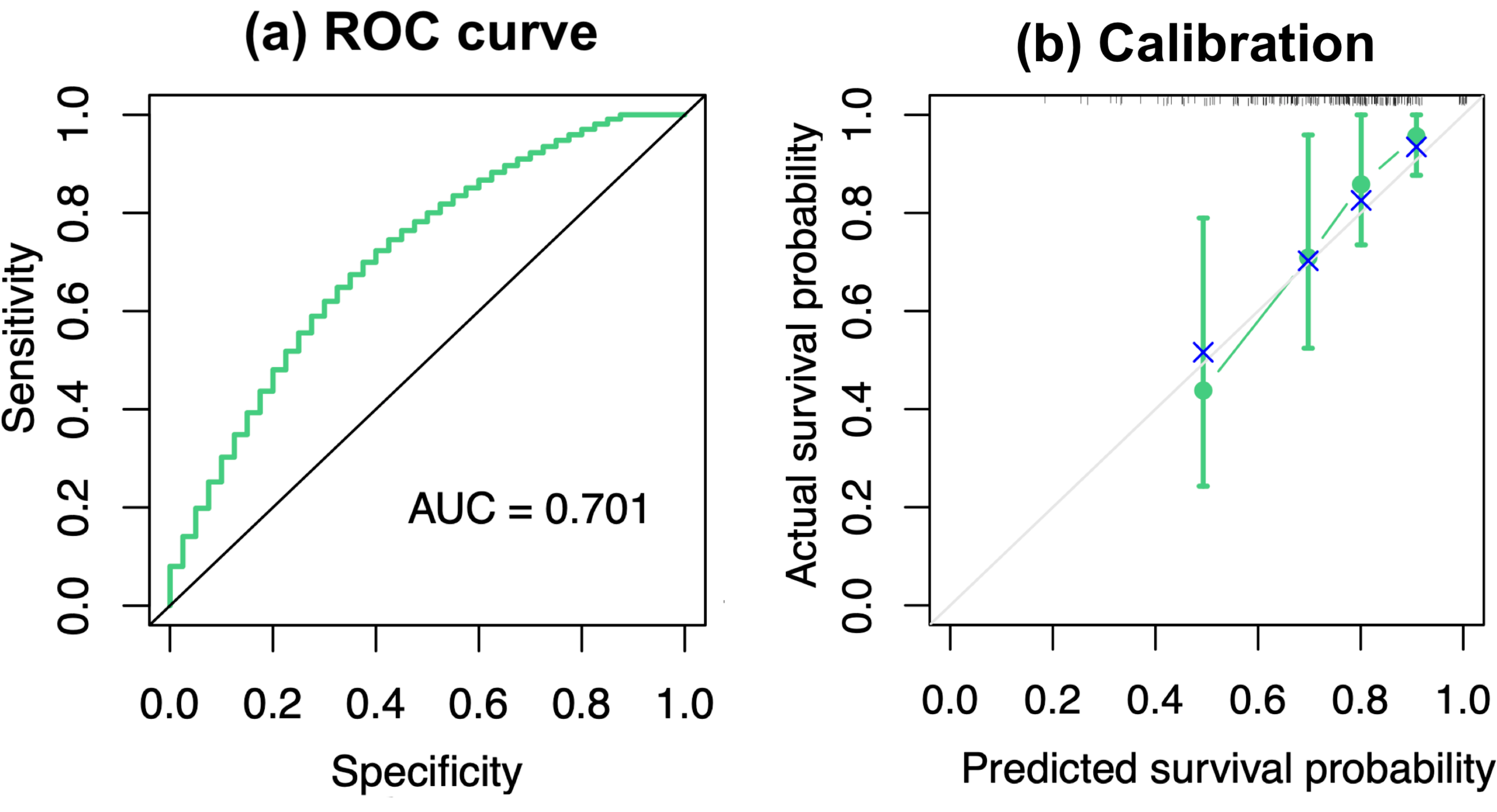}
  \caption{\it Discrimination and calibration of survival prediction for 20\% TCGA-BRCA validation data. TCGA-BRCA data were split to a 80\% training set and a 20\% validation set. (a) Receiver operating characteristic (ROC) curve estimated at 5-year survival evaluation time point based on the Cox model in \hyperref[fig:nomogram]{Figure~\ref*{fig:nomogram}}. The AUC score is the area under the ROC curve. The 45-degree line represents the performance of a random prediction of the outcome event with AUC = 0.5. (b) Calibration plot estimating 5-year survival probabilities in TCGA-BRCA patients grouped by the quartiles of the predicted probabilities based on the Cox model in \hyperref[fig:nomogram]{Figure~\ref*{fig:nomogram}}. The actual 5-year survival probabilities in each group (green colored dot and error bar) were estimated by the Kaplan-Meier (KM) method with a 95\% confidence interval by bootstrapping. The blue colored cross indicates a bias-corrected KM estimate. The predicted survival probability of each quartile group was the mean of the 5-year survival probabilities based on the Cox model for the corresponding group.}\label{fig:calibration}
\end{figure}

When an independent validation data set is available, it is recommended to report a calibration plot corresponding to the nomogram. 
Using independent validation data to obtain $S_{\text{model}}(t|g)$ in the calibration model (\ref{eq:calibration}) is for the generalization capacity of the model. 
Since we here do not have independent validation data besides TCGA-BRCA data, \hyperref[fig:calibration]{Figure~\ref*{fig:calibration}} shows an example calibration plot at 5-year survival evaluation time point based on the built Cox model in \hyperref[fig:nomogram]{Figure~\ref*{fig:nomogram}} for a split 20\% TCGA-BRCA data set. 

\section{Beyond penalized and Bayesian Cox models}

In this tutorial, we mainly focused on penalized regressions and Bayesian hierarchical models in the Cox proportional hazards framework. 
One can extend this framework in several ways. 
For instance, one can stay in a likelihood-based modeling framework, but replace the partial likelihood function of the semi-parametric Cox model by alternative likelihood functions (which do not necessarily need to imply proportional hazards), e.g., parametric survival models like exponential, Weibull, or accelerated failure time (AFT) models, or Aalen's additive hazard model \citep{GorstRasmussen2012}. 
Alternatively, one can move to a more algorithmic machine learning approach, such as tree-based boosting or bagging methods, e.g., random survival forests \citep{Hothorn2006,Binder2008,Jaeger2019}, 
or (deep) neural networks \citep{Wiegrebe2023}.

\cite{Hothorn2006} introduced ensemble tree methods for analyzing right-censored survival data, which construct ensembles from base learners, e.g., binary survival trees for each omics feature. 
\cite{Hothorn2006} also proposed a gradient boosting algorithm to predict the survival time of patients with acute myeloid leukemia (AML), based on clinical and omics features. 
Similarly, \cite{Binder2008} developed a likelihood-based boosting method, which aims to maximize the Cox partial likelihood function, for modeling time-to-event data based on high-dimensional omics input data and which also allows the inclusion of a small number of mandatory covariates. 
In general, one needs to be cautious if using some machine learning methods that are not well-suited for high-dimensional features. For example, \cite{Kvamme2019} proposed extensions of the Cox model with neural networks, which are only valid if the number of covariates is smaller than the number of samples, i.e., if $p < n$. 
A systematic review of deep learning for survival analysis, which includes a survey of methods suitable for high-dimensional data ($p > n$), is provided by \cite{Wiegrebe2023}. 

In the case of non-proportional hazards, many likelihood-based survival models beyond Cox-type models have also been extended to account for high-dimensional omics as input data. 
For example, \cite{Ma2006} combined Lin and Yin's additive hazard model \citep{Lin1994} with principal component regression for dimension reduction of omics features, which was applied to the study of gene expression-based survival prediction for diffuse large B-cell lymphoma. 
\cite{Engler2009} added the Elastic Net penalty in an accelerated ATF model, which assumes that the effect of a covariate accelerates or decelerates the life course of patients. 
\cite{Schmid2008} and \cite{Barnwal2022} used boosting algorithms to learn parametric AFT models. 
\cite{Ha2014} considered the Lasso, SCAD and a penalized h-likelihood for feature selection in frailty models which assume that individuals have unobserved heterogeneity captured by a latent random term $Z$, which adapts the Cox model (\ref{eq:Cox}) into $h(t|X) = Zh_0(t)\exp(X\beta)$. 

\subsubsection{Advanced survival models: cure models, competing risks and multi-state models}

In some situations, survival data may be different from \hyperref[fig:omics]{Figure~\ref*{fig:omics}c} (also Section ``\hyperref[sec:datasurvival]{Time-to-event data}''), where it was presumed that all individuals will eventually experience the event of interest. 
\cite{Liu2012} studied the Lasso and SCAD feature selection for the proportional hazard mixture cure model, in which a certain fraction of individuals will never experience the event of interest. 
\cite{Tapak2015} investigated Lasso, Elastic Net and likelihood-based boosting for microarray-based survival modeling with competing risks, such as ``progression'' versus ``death from non-cancer cause'', i.e., the event of a patient can occur due to one of multiple distinct causes. 
There is a growing awareness of the impact of competing risks when developing prognostic models with high-dimensional input data, for example, \cite{Binder2009}, \cite{Ambrogi2016} and \cite{Fu2017}. 
For a single individual who can experience several possible events, \cite{Dutta2016} proposed a multi-state model to identify risk factors in different stages of disease based on high-dimensional input data.

\section{Towards single-cell data analysis}\label{sec:single-cell}

The cellular heterogeneity of complex sample mixtures pose challenges and also opportunities for precision medicine and survival prediction. 
For example, \cite{Zhou2019} showed that tumor microenvironment-related gene expression signatures do not only accurately predict the survival among colon cancer patients, but also serve as biomarkers for identifying patients who could potentially benefit from adjuvant chemotherapy. 
Single-cell technologies provide an unprecedented opportunity for dissecting the interplay between the cancer cells and the associated tumor microenvironment, and the produced high-dimensional data should also augment existing survival modeling approaches. 
The emerging single-cell atlases are providing a detailed and quantitative overview of tissue composition and organization, and will advance both biomedical research and clinical practice \citep{Elmentaite2022}. 

Currently, the focus of statistical model development for single-cell data analysis is to understand cell type composition and its impact on gene regulation and transcriptional dynamics, usually based on only a small number of samples/individuals. 
On the one hand, the underlying statistical models for single-cell data analysis are still in development and continuously being re-evaluated and challenged \citep{Kharchenko2021}. 
On the other hand, survival analysis tackles disease and health contexts at the individual level, and usually requires a relatively large number of individuals for sufficient statistical power. 
In particular, the development of improved computational methods is urgently needed to enable the consideration of multiple layers (e.g., individual-level, cellular-level, molecular-level) of information when integrating groups of individuals and omics data from a variety of molecular modalities. 
Therefore, current approaches often map the findings from single-cell omics data to large-scale bulk sequencing omics and survival data, rather than jointly analyzing single-cell omics and survival data. 
For example, \cite{Guo2018} introduced a generalizable approach to first study T-cells in 14 non-small cell lung cancer (NSCLC) patients and to identify gene signatures from the tumor-enriched T cell clusters, and then investigated these gene signatures using bulk RNA-seq and survival data from larger TCGA-NSCLC patient cohort. 
Similarly, \cite{Zhang2020} developed a scRNA-seq-based approach to reconstruct a multilayer signaling network based on 16 glioma patients, and then investigated the network genes using survival data from TCGA Chinese glioma genome atlas (TCGA-CGGA) patients. 
However, direct joint analysis of survival and single-cell omics data from multiple cellular hierarchies requires further methodological developments and new statistical and machine learning methods. 

\section{Discussion}

Although survival analysis faces many modeling challenges, mainly due to censored outcomes, it represents a well-established methodology for finding risk factors associated with patients' survival. 
The identification of omics biomarkers for survival prognosis may provide systematic means to guide patient management and personalized treatment and diagnostic strategies. 
In this tutorial, we provided a comprehensive workflow for survival analysis with high-dimensional omics and standard clinical data, with a specific focus on feature selection of survival-associated omics features and survival model validation. 
We covered many penalized regressions and Bayesian models for feature selection and survival prediction, accounting for their specific assumptions and applications. 
Examples of real data and R scripts have been made available to illustrate the use of different methods, which should help researchers to choose and apply suitable methods for their survival analysis applications (\url{https://ocbe-uio.github.io/survomics/survomics.html}). 
We note that this review only considers methods for right-censored time-to-event data, i.e., where all individuals are assumed to be followed continuously from the start of the study, but where the follow-up period might end before the event (e.g., death) was observed. 
Other types of censoring include interval censoring and left truncation, and appropriate statistical methods dealing with these censoring patterns should be chosen accordingly. 

Most of the current methods for survival analysis do not explicitly take into account the complex structures within and between multi-omics data, such as gene regulation and DNA-protein interactions. 
Regulatory networks constructed either based on prior biological knowledge or using data-driven, yet biologically explainable approaches, may help establish useful methodologies for survival analysis that are more effective for deriving biological insights as well as enable improved clinical translation. 
However, to achieve a comprehensive and biologically meaningful integration of high-dimensional multi-omics data, there is a need for continued development of computational and statistical approaches that consider both technical and biological intricacies of the data and technologies, respectively \citep{Wissel2023}. 
This is currently a very active research field, and we expect to see many improved multi-omics methods for survival prediction in the future.

Another limitation of most of the reviewed methods is that they identify omics features prognostic of survival, but they cannot determine causal relationships. 
Causality is a fundamental notion to understand omics features causing disease progression, which will allow one to reliably intervene omics features for targeted therapies. 
There are two popular causal inference models, Pearl's structural causal model (SCM) and Rubin's causal model (RCM), both of which introduce perturbations to draw causal inference. 
\cite{Farooq2023} utilized SCM-based causal discovery approaches to unravel relationships between omics features and survival of breast cancer patients. 
However, to identify reliable causal relations for clinical applications, laboratory-based experiments, e.g., clustered regularly interspaced short palindromic repeats (CRISPR) techniques \citep{Wang2023}, are often necessary to verify the functional relevance of the identified omics features.  
High-dimensional RCM-based mediation analysis has been used to investigate the indirect effect transmitted by omics features between an exposure and survival outcomes \citep{Song2020,Song2021}. 
Causal mediation analysis is an important tool, which considers the problem of decomposing the causal effect of treatment/exposure into direct and indirect effects \citep{Lange2011,VanderWeele2011}. 
The direct effect corresponds to the effect of a treatment directly on the survival outcome, while an indirect effect corresponds to the effect of a treatment on the outcome that is due to its effect on an intermediate variable (e.g., gene expression) that also has a causal effect on the survival outcome.
Targeted learning also fills a much needed gap between statistical modeling and causal inference \citep{vanderLaan2011,vanderLaan2018}. 
\cite{Tuglus2011} used targeted maximum likelihood estimation to provide an interpretable causal measure of variable importance for the discovery of biomarkers and omics features. 
Another way to formalize personalized medicine is dynamic treatment regimes \citep{Chakraborty2013,Tsiatis2019,Deliu2023} that encompasses causal inference and takes into account for the variability in omics, environment and lifestyle factors for each individual to improve the treatment of a particular patient. However, all the causal machine learning methods require further methodological developments for adaptation to survival modeling with high-dimensional input data.

\section*{Data availability}

Supplementary step-by-step R tutorial is available online at \url{https://ocbe-uio.github.io/survomics/survomics.html}. TCGA data is publicly available at \url{https://portal.gdc.cancer.gov}.

\section*{Conflict of interest}

None declared.

\section*{Acknowledgements}

The authors thank Professor Ørnulf Borgan for his valuable suggestions and comments. This work was supported by grants from Helse S\o{}r-\O{}st (grant 2020026 to TA), the Norwegian Cancer Society (216104 to TA), the Radium Hospital Foundation, the Academy of Finland (grants 326238, 340141, 344698, 345803 to TA), Cancer Society of Finland (to TA), the European Union’s Horizon 2020 research and innovation programme (grant `PANCAIM' 101016851 to JZ and TA, grant `RESCUER' 847912 to MZ), the Innovative Medicines Initiative 2 Joint Undertaking of the European Union’s Horizon 2020 research and innovation programme and EFPIA and JDRF INTERNATIONAL (grant `imSAVAR' 853988 to MZ), ERA PerMed under the ERA-NET Cofund scheme of the European Union’s Horizon 2020 research and innovation framework programme (grant `SYMMETRY' 342752 to MZ).

%USE THE BELOW OPTIONS IN CASE YOU NEED AUTHOR YEAR FORMAT.

\bibliographystyle{apalike}
\bibliography{reference}

\end{document}